\documentclass[fleqn,usenatbib]{mnras}

\usepackage{newtxtext,newtxmath}
\usepackage[T1]{fontenc}
\usepackage[utf8]{inputenc}
\usepackage{ae,aecompl}
\usepackage{multirow}
\usepackage[normalem]{ulem}

\usepackage{graphicx}	% Including figure files
\usepackage{amsmath}	% Advanced maths commands
\usepackage[usenames, dvipsnames]{color}

\usepackage{hyperref}
\newcommand{\msun}{\mbox{$\rm M_\odot$}}

\newcommand{\hagn}{\mbox{{\sc \small Horizon-AGN}}}

\newcommand{\nh}{\mbox{{\sc \small NewHorizon}}}
\newcommand{\hs}{\mbox{{\sc \small Horizon}}}

\title[IMBH in dwarf galaxies]{Population statistics of intermediate mass black holes in dwarf galaxies using the NewHorizon simulation}

\author[Beckmann, R.S.]{Beckmann, R.S.$^{1}$\thanks{E-mail: ricarda.beckmann@ast.cam.ac.uk },
Dubois, Y.$^{2}$,
Volonteri, M.$^{2}$,
Dong-P\'aez, C. A.$^{2}$, \newauthor 
Trebitsch, M.$^{3}$,
Devriendt, J.$^{4}$,
Kaviraj, S.$^{5}$,
Kimm, T.$^{6}$,
Peirani, S.$^{7}$
\\
$^{1}$Institute of Astronomy and Kavli Institute for Cosmology, University of Cambridge, Madingley Road, Cambridge, CB3 0HA, UK\\
$^{2}$Institut d’Astrophysique de Paris, CNRS, Sorbonne Universit\'e, UMR7095, 98bis bd Arago, 75014 Paris, France\\
$^{3}$Kapteyn Astronomical Institute, University of Groningen, P.O. Box 800, 9700 AV Groningen, The Netherlands \\
$^{4}$ University of Oxford, Astrophysics, Denys Wilkinson Building, Keble Road, Oxford OX1 3RH, UK \\
$^{5}$ Centre for Astrophysics Research, Department of Physics, Astronomy and Mathematics, University of Hertfordshire, Hatfield, AL10 9AB, UK \\
$^{6}$ Department of Astronomy, Yonsei University, 50 Yonsei-ro, Seodaemun-gu, Seoul 03722, Republic of Korea \\
$^{7}$Universit\'e C\^ote d'Azur, Observatoire de la C\^ote d'Azur, CNRS, Laboratoire Lagrange, Bd de l'Observatoire,CS 34229, \\ 
06304 Nice Cedex 4, France \\
}

\date{Accepted XXX. Received YYY; in original form ZZZ}

\pubyear{2015}

\begin{document}
\label{firstpage}
\pagerange{\pageref{firstpage}--\pageref{lastpage}}
\maketitle

% Abstract of the paper
\begin{abstract}
While it is well established that supermassive black holes (SMBHs) co-evolve with their host galaxy, it is currently less clear how lower mass black holes, so-called intermediate mass black holes (IMBHs), evolve within their dwarf galaxy hosts. In this paper, we present results on the evolution of a large sample of IMBHs from the \nh~ simulation. We show that occupation fractions of IMBHs in dwarf galaxies are at least 50 percent for galaxies with stellar masses down to $10^6 \rm \ M_\odot$, but BH growth is very limited in dwarf galaxies. In \nh, IMBH growth is somewhat more efficient at high redshift $z=3$ but in general IMBH do not grow significantly until their host galaxy leaves the dwarf regime. As a result \nh~ under-predicts observed AGN luminosity function and AGN fractions. We show that the difficulties of IMBH to remain attached to the centres of their host galaxies plays an important role in limiting their mass growth, and that this dynamic evolution away from galactic centres becomes stronger at lower redshift.
\end{abstract}

% Select between one and six entries from the list of approved keywords.
% Don't make up new ones.

%%%%%%%%%%%%%%%%%%%%%%%%%%%%%%%%%%%%%%%%%%%%%%%%%%

%%%%%%%%%%%%%%%%% BODY OF PAPER %%%%%%%%%%%%%%%%%%

\section{Introduction}

Supermassive black holes (SMBHs) with masses of $M_{\rm BH}>10^{7} \msun$ or above, are well known to tightly correlate with properties of their host galaxy, such as the stellar bulge mass and the stellar velocity dispersion \citep[see][ for a review.]{Kormendy2013}. While these correlations are firmly established observationally for SMBHs, little data is available for black holes (BHs) in the intermediate BH mass range $10^{4} < M_{BH} < 10^6 \msun$, so called intermediate mass black holes (IMBHs, in the following the acronym `BHs' refers to black holes of all masses). If extrapolations of the correlations into the intermediate mass regime hold, such IMBHs could play the same role in dwarf galaxies that SMBHs play in massive galaxies. 

IMBHs, and their potential role in shaping dwarf galaxies, is less well understood from both a simulation and an observation point of view.  Observationally, the inherently lower luminosity of IMBHs due to their low BH masses make them difficult to detect and to distinguish from star formation \citep[see][for a review]{Greene2020}. From a theoretical point of view, the high resolution required to resolve the internal structure of simulated dwarf galaxies means they are often unresolved in large scale galaxy evolution simulations, which can therefore not be used to study the coevolution of IMBHs and their host galaxies \citep{Haidar2022}. Instead, specific simulations targeting the dwarf galaxy regime are required to study such galaxies and their central BHs.

Over the last decade, great progress has been made  to expand the sample of observed IMBHs, as has been summarised in the recent reviews by \citet{Greene2020} and \citet{Mezcua2017}. There are two main methods for detecting IMBHs: kinematical studies detect BHs via the impact of their gravitational potential on host galaxy stars. This technique makes it possible to detect quiescent BHs and to directly measure the BH mass, but is limited to nearby galaxies. Alternatively, IMBHs can be detected when accreting as active galactic nuclei (AGN) in a variety of wavelengths: in the optical \citep{Molina2021,Mezcua2018,Mezcua2020,Manzano-King2020}, the X-ray \citep{Chilingarian2018,Latimer2021,Toptun2022}, the radio \citep{Davis2022,Yang2020,Reines2019} or the infrared \citep{2020MNRAS.492.2528L} to highlight a few recent studies. Other recent detection methods being explored are  gamma-ray bursts potentially lensed by an IMBH \citep{Paynter2021}, short-term variability \citep{Shin2021}. In the future, gravitational wave observations will provide another opportunity to detect IMBHs \citep{2005ApJ...623...23S,Ricarte2018,Bellovary2018,2018PhRvD..98f3018A,2020MNRAS.491.2301K,2021MNRAS.500.4095V,DeCun2021,2011GReGr..43..485G}. The sample of IMBHs in dwarf galaxies is therefore very much still in the process of being assembled, and it is expected to expand significantly in coming years.

 Observational surveys to date have shown that AGN fractions in dwarf galaxies are in the typically observed to be in the range of $0.1-5$ percent \citep{Latimer2021,Mezcua2020,Wylezalek2017,Reines2013, Pardo2016,Aird2018,Birchall2020} but may be as high as 30 percent \citep{2019MNRAS.489L..12K,Dickey2019,Davis2022} and depend strongly on the AGN selection method \citep{Mezcua2020,2020MNRAS.492.2528L,Greene2020} and the observational proxy for BH mass \citep{Gallo2019} chosen. Firm IMBH mass detections are only available for a small sample of objects, and even fewer also have kinematics measurements of the host galaxy. For those that have both, evidence is mounting that the M$_{\rm BH}-\sigma$ relation shows no evidence of a break in slope in the dwarf galaxy regime \citep{Greene2020,Baldassare2020,Nguyen2019}. This lack of break has been argued to be a sign that IMBHs in dwarf galaxies regulate the evolution of their host galaxy through feedback in the same way that massive galaxies are regulated by SMBHs \citep{King2021}. Further evidence for the theory of IMBH-regulated dwarf galaxies comes from studies of gas kinematics in dwarf that show significantly more disturbed morphologies for those with AGN than those without \citep{Manzano-King2020}.

In simulations, our current picture of the co-evolution between BH and host galaxy is more mixed: some show a break in the correlations around the transition from dwarf galaxy to massive galaxy \citep{Sharma2019,Koudmani2021}, while others merely find an increase in scatter and no break \citep{Ricarte2019,Barai2019,Sharma2022}.  As many groups find that the growth of BHs in low-mass galaxies is regulated by supernova (SN) feedback \citep{Dubois2015,Habouzit2017,Bower2017,Angles-Alcazar2017,Trebitsch2018}, whether BHs are over-massive or under-massive in comparison to the correlation depends strongly on how both supernova feedback and BH accretion and feedback are modelled \citep{Koudmani2022}. With current SN feedback models, BH growth in dwarf galaxies is restricted mostly to high redshift \citep{Barai2019,Koudmani2021} but recent evidence from simulations suggests AGN feedback in dwarf galaxies might continue to impact both star formation and galactic outflows with with strong SN feedback \citep{Koudmani2019,Nelson2019} and that a wider variety of feedback models can lead to AGN feedback playing an even more important role in the evolution of dwarf galaxies \citep{Koudmani2022}. Analytic models also argue that the fraction of active BHs in dwarf galaxies could be higher than currenlty observed in X-ray samples \citep{Pacucci2021}. On the other hand, widespread growth of IMBHs in dwarf galaxies would lead to an overproduction of faint AGN in tension with the observed AGN luminosity function \citep{Habouzit2017,2022MNRAS.511.5756T}.

Both observations \citep{Mezcua2020} and simulations \citep{Bellovary2018,Bellovary2021} frequently show IMBHs in galaxies that are not located at the center of the galaxy. The discussion around the reason for this phenomenon is ongoing, but is likely linked to the evolution history of the host galaxy. \citet{Boldrini2020} use isolated halos to show that DM subhalos falling onto dwarf galaxies can displace the central IMBH by 100 pc or more, while \citet{Bellovary2021} show that the off-centre location in their sample of dwarf galaxies in a cosmological environment is due to mergers. While there generally is a large `hidden' population of BHs in dwarf galaxies which are not sufficient active to be observable \citep{2009MNRAS.400.1911V}, the percentage of hidden BHs that are off-centre is particularly high \citep{Sharma2022}.

In this paper, we present the sample of IMBHs in dwarf galaxies, which are here defined to have a mass of $M_{\rm dwarf} = 3 \times 10^9 \rm \ M_\odot$, in the \nh~ simulation\footnote{https://new.horizon-simulation.org/} \citep{Dubois2021}. \nh~ is a cosmological zoom simulation of an average density volume of the Universe that has sufficiently high resolution to resolve galaxies down to a stellar mass of $10^6 \msun$. We use this  sample to study correlations and population statistics of IMBHs and their dwarf galaxy hosts to expand our understanding of the coevolution of massive galaxies and their BHs into the dwarf galaxy regime. The paper is structured as follows: the simulation setup is briefly recapped in Section \ref{sec:simulation}. BH mass evolution is discussed in Section \ref{sec:scaling-relations}, BH occupation ratios are shown in Section \ref{sec:occupation} and mass functions are discussed in Section \ref{sec:mass_function}. The detectability of BHs is discussed in Sections \ref{sec:bolometric_luminosity} and \ref{sec:AGN_frac}, while the distribution of IMBHs within their host galaxies is analysed in Section \ref{sec:orbits}. Section \ref{sec:conclusions} summarises the paper.

\section{Simulation}
\label{sec:simulation}
\nh\ is a high-resolution resimulation of an average spherical sub-volume with a radius of 10 comoving Mpc of the \hagn\ simulation \citep{Dubois2014}. \nh\ has been presented in detail in \citep{Dubois2021}. \nh~was run until $z=0.25$.

The simulation assumes a $\Lambda$CDM cosmology consistent with WMAP-7 data \citep{komatsu2011} with a dark energy density $\Omega_{\Lambda}=0.728$, baryon density $ \Omega_{\rm b}=0.045$, total matter density $\Omega_{\rm m}=0.272$, a Hubble constant of $H_{0}=70.4 \rm \ km\,s^{-1}\, mpc^{-1}$, and an amplitude of the matter power spectrum and power-law index of the primordial power spectrum of $\sigma_8=0.81$ and $n_{\rm s}=0.967$ respectively. A high-resolution region of radius of 10 comoving Mpc with a DM mass resolution of $ 1.2 \times 10^6 \msun$ is embedded  within the 142 a side comoving Mpc box of \hagn.

All simulations within the \hs\ suite  were performed using {\sc ramses} \citep{Teyssier2002}, using a second-order unsplit Godunov scheme for solving the Euler equations, and an HLLC Riemann solver with a MinMod Total Variation Diminishing  scheme  to  reconstruct  interpolated  variables. In \nh\, refinement proceeds according to a quasi-Lagrangian scheme up to a maximum resolution of $34 \rm pc$ at z=0, where a cell is refined if its mass exceeds 8 times the initial mass resolution. The minimum cell size is kept approximately constant throughout by adding an extra level of refinement at expansion factor $a_{\rm exp}=0.1,0.2,0.4$ and $0.8$. This is supplemented with a super-Lagrangian refinement criterion that enforces refinement of cells whose size is smaller than one Jeans length wherever the gas number density is larger than $5\, \rm H\, cm^{-3}$.

The gas follows an equation of state for an ideal monoatomic gas with an adiabatic index of $\gamma_{\rm ad}=5/3$. Gas cooling is modelled using cooling curves from \cite{Sutherland1993} down to $10^4\, \rm K$, assuming equilibrium chemistry.  Heating from a uniform UV background takes place after redshift $z_{\rm reion} = 10$ following \cite{Haardt1996}. 

Star formation occurs in cells whose hydrogen gas number density exceeds $n_0=10\, \rm H\, cm^{-3}$ following a thermo-turbulent sub-grid algorithm in combination with a Schmidt law \citep{kimmetal17,trebitschetal17,trebitschetal20}. The stellar mass resolution is $ 1.3\times 10^4 \ \msun$), and stars are assumed to have a Chabrier \citep{Chabrier2005} initial mass function with cutoffs at 0.1 and 150 \msun. Stellar feedback is modelled following \citet{Kimm2015}, which  separately tracks the momentum and energy conserving phase of the explosion, which reproduces the stellar-mass-to-halo-mass relation in dwarfs \citep{Dubois2021}.

%BH formation, merger and dynamics
New BHs are formed in any cell in which the gas and stellar density exceeds the threshold for star formation, which has a stellar velocity dispersion of more than 20 $\rm km s^{-1}$ and that is located more than 50 kpc from any existing BH. Each BH is formed with a mass of $M_{\rm BH,0}=10^4 \ \msun$ and an initial spin of $a=0$. We note that this is slightly below the stellar mass resolution of $1.3 \times 10^4 \rm \ M_\odot$ which might have important consequences for how well the dynamics of BH near their seed mass is resolved in \nh. To avoid spurious motions of BHs due to finite force resolution effects, we include an explicit drag force of the gas onto the BH, following \cite{Ostriker1999}. Two BH particles are allowed to merge into a single BH particle when they get closer than $4\Delta x$ ($\sim 150$~pc) and when the relative velocity of the pair is smaller than the escape velocity of the binary. A detailed analysis of BH mergers in \nh\ is presented in~\cite{Volonteri2020}. 

%Growth and feedback
BHs grow through un-boosted Bondi-Hoyle-Lyttleton accretion 
\begin{equation}
\dot{M}_{\rm BHL} =\frac{4 \pi (G M_{\rm BH})^2 \rho}{(c_{\rm s}^2+v_{\rm rel}^2)^{3/2}}
\label{eq:dotMbhl}
\end{equation} where $\rho$, $c_{\rm s}$ and $v_{\rm rel}$ are the local mass-weighted, kernel weighted gas density, sound speed and relative velocity between gas and BH. The accretion rate is capped at the BH's Eddington rate $\dot{M}_{\rm Edd}$, which is calculated using the spin-dependent radiative efficiency $\varepsilon_{\rm r}$: 
\begin{equation}
    \varepsilon_{\rm r}=f_{\rm att}\left(1-E_{\rm isco} \right)=f_{\rm att}\left(1-\sqrt{1-2/(3r_{\rm isco})}\right)
\label{eq:epsilonr_spin}
\end{equation}
where $r_{\rm isco}=R_{\rm isco}/R_{\rm g}$ is the radius of the innermost stable circular orbit (ISCO) in reduced units and $R_{\rm g}$ is half the Schwarzschild radius of the BH. $R_{\rm isco}$ depends on spin $a$.
For the radio mode, the radiative efficiency used in the effective growth of the BH is attenuated by a factor $f_{\rm att}=\min(\chi/\chi_{\rm trans},1)$  following~\cite{benson&babul09}, where $\chi$ is the Eddington ratio.

During accretion, a fraction of the accreted energy is returned to the gas in one of two AGN feedback modes:  1. A quasar mode at Eddington ratios $\chi > 0.01$, in which energy is injected  isotropically around the BH as thermal energy using a fixed efficiency of 15\%  2. A jet mode at Eddington ratios $\chi < 0.01$, in which energy is injected as kinetic energy in bipolar jets aligned with the BH's spin axis using a spin-dependent efficiency that is higher for high spin values \citep[see][for details]{Dubois2021}.

%spin evolution
BH spin is followed on-the-fly in the simulation, taking into account the angular momentum of accreted gas, BH-BH mergers and BH spindown during jet mode feedback. The BH spin model is presented in detail in \citet{Dubois2014spin}. When the BH is accreting with an Eddington fraction $\chi > 0.01$, BHs are either spun up or down depending on whether the angular momentum of the accreted gas feeds an aligned or misaligned sub-grid disc \citep{kingetal05}. For accretion at lower luminosity, the BH-driven jets are  assumed to be powered by energy extraction from BH rotation \citep{Blandford1977}, and as a consequence  the BH spin magnitude can only decrease. During mergers, the spin of the remnant is calculated according to the spin of the initial BHs, and the angular momentum of the binary, according to \citet{rezzollaetal08}.

All physics included in \nh\ are described in further detail in \citep{Dubois2021}. The spin evolution of black holes in \nh~ is analysed in a companion publication \citep{Beckmann2022d}.

\subsection{Black hole and galaxy catalogue}
\label{sec:catalogue}

\begin{figure*}
    \centering
    \includegraphics[width=\textwidth]{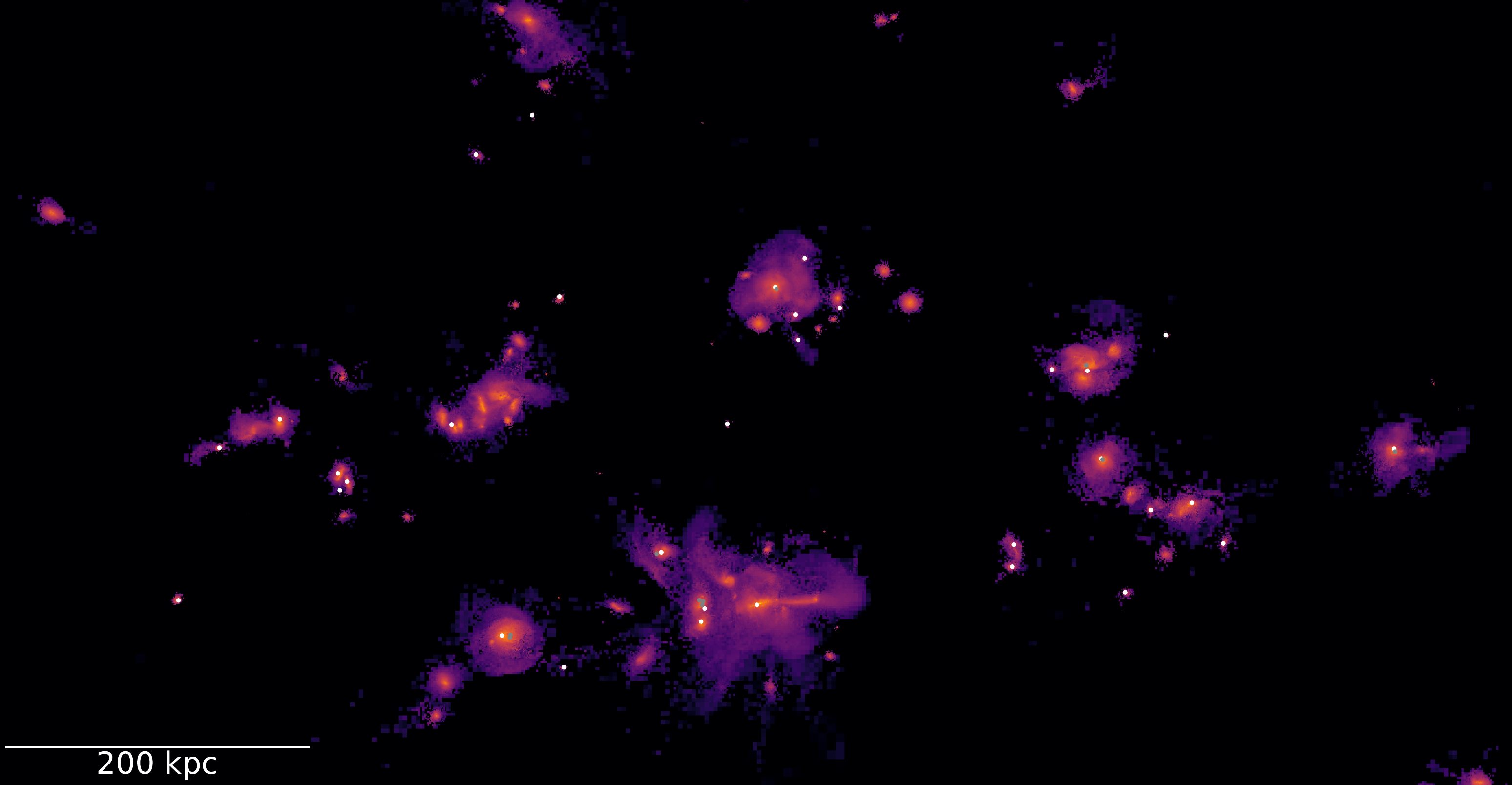}
    \caption{Stellar density projection of the central region of \nh~ at $z=2$. Main BHs are overplotted in white, while secondary BHs are shown in grey.}
    \label{fig:pretty_picture}
\end{figure*}

The sample of BHs discussed in this paper includes all BHs contained within \nh\ that are associated with a host galaxy from the galaxy catalogue, which in turn has to be associated with a host halo from the halo catalogue.

The DM halo catalogue consists of all un-contaminated DM halos identified by the structure-finding algorithm HOP \citet{Eisenstein1998}. A halo is considered uncontaminated if all DM particles contained within it originate from the zoom region, i.e. are at the maximum DM mass resolution. 

Galaxies are identified using HOP applied to the star particles of the simulation. The galaxy consists of all galaxies flagged as level 1 (i.e. main or central structure for a given DM halo), or that have a stellar mass above $10^8 \msun$ if they are satellites. Galaxies are only included in the galaxy catalogue if they are located within the central 0.1 virial radii of a halo contained in the halo catalogue. The centres of galaxies and halos are identified using an iterative shrinking-sphere approach \citep{Power2003}.

To identify BHs with their host galaxy, we cycle over all galaxies in the catalogue from most to least massive, and for each galaxy identify the most massive BH to be contained within two effective radii of the galaxy's centre. Galaxy effective radii are defined to be the geometric mean of the half-mass radius of the projected stellar densities along each of the Cartesian axis. This BH is flagged as the galaxy's main BH and removed from the sample of unallocated galaxies. We then repeat the loop over all galaxies from most to least massive, identifying all as-of-yet unallocated BHs contained within 2 effective radii of the galaxy as secondary BHs of that galaxy. A galaxy can contain multiple BHs, but a BH can only be associated with a single galaxy. The full sample of BHs discussed in this paper contains all BHs associated with host galaxies. BHs contained in contaminated galaxies and halos, and ``wandering" BHs that are far from the luminous part of any galaxy are excluded from the sample. The distribution of main BH in and around galaxies for the central region of the box at $z=2$ can be seen in Fig. \ref{fig:pretty_picture}.

\section{Results}

%Based on http://localhost:8888/notebooks/2020_05_07_newHbh_correlation_relations.ipynb

\subsection{Black hole - galaxy correlations}
\label{sec:scaling-relations}

\begin{figure}
    \centering
    \includegraphics[width=\columnwidth]{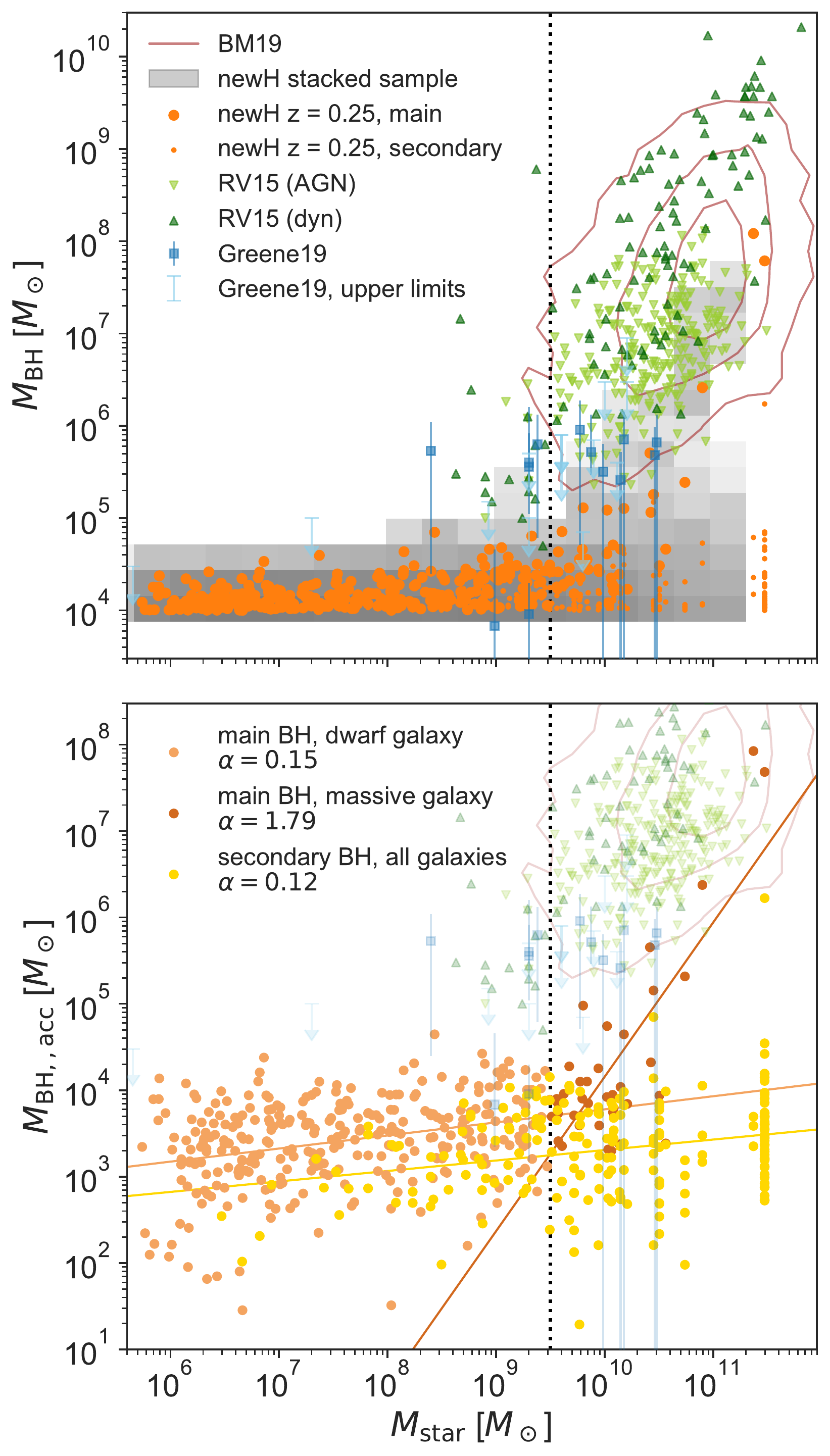}
    \caption{Correlations between mass  and host galaxy properties at $z=0.25$: [Top] Galaxy stellar mass versus BH mass $M_{\rm BH}$. The grey distribution shows that stacked sample at all redshifts.  [bottom] Galaxy stellar mass versus accereted BH mass $M_{\rm BH,acc}$. Also shown for comparison are observational data from \citet{reines&volonteri15} (RV15, green triangles), \citet{baron&menard2019} (BM19, grey contours). Fits from \citet{Kormendy2013} are shown in black, and fits from \citet{Greene2020} in blue. The same observations are shown on both panels. Errorbars in RV15 are omitted for clarity. Galaxies left of the dotted black line are considered dwarf galaxies.}
    \label{fig:MBH_Mstar}
\end{figure}

As can be seen in the top panel of Fig. \ref{fig:MBH_Mstar}, BHs in dwarf galaxies ($M_{\rm star} < M_{\rm dwarf} = 3 \times 10^9 \rm \ M_\odot$) in \nh\ grow little beyond their seed mass of $10^{4} \rm \ M_\odot$ by $z=0.25$.  It is only once their host galaxy leaves the dwarf regime that main BHs grow efficiently onto the observed $M_{\rm star}-M_{\rm BH}$ correlation. Secondary BHs continue to struggle to grow beyond $10^5 \rm \ M_\odot$. As a result, the $M_{\rm BH}-M_{\rm star}$ relation shows a clear break in slope around $M_{\rm dwarf}$.

In the bottom panel of Fig. \ref{fig:MBH_Mstar}, we study in more detail how efficient BHs in dwarf galaxies grow over time by plotting $M_{\rm BH, acc}$, the \emph{accreted} mass. $M_{\rm BH, acc}$ is the mass of each BH gained through gas accretion alone. It is found by taking the fiducial BH mass, $M_{\rm BH}$ and subtracting both the seed mass $M_{\rm BH,0}=10^4 \rm \ M_\odot$ as well as any mass gained through BH-BH mergers. As can be seen by the slope $\alpha$ of the linear fit to $M_{\rm BH, acc}$ vs $M_{\rm star}$ (solid lines), BHs in more massive dwarf galaxies do grow somewhat more than in low-mass dwarfs, but the slope of this trend is much shallower than for main BHs in massive galaxies. It is, however, very similar to that for secondary BHs in galaxies of all stellar masses. We caution that while the BH seed mass $M_{\rm BH,0}$ is no longer included in $M_{\rm BH, acc}$, the distribution of $M_{\rm BH, acc}$ still depend on $M_{\rm BH,0}$ as the BH accretion rate depends explicitly on the instantaneous BH mass at any given point in time (see Eq.\ref{eq:dotMbhl}). If we had decided to set $M_{\rm BH,0}$ to a higher value initially, our BH would have grown faster than a BH with lower $M_{\rm BH,0}$ in the same environment.
 
This change in slope is different to tentative observational conclusions, which show no evidence so far for a break in slope in the $M_{\rm BH}-M_{\rm star}$ relation during the transition from dwarf to massive galaxy \cite[see][for a review]{Greene2020}, and observational data points plotted for comparison in Fig. \ref{fig:MBH_Mstar}. The lack of change in slope could be due to an observational bias, as over-massive BHs are easier to observe for a given galaxy stellar mass than under-massive ones. No such bias exists in simulations, which could potentially skew the comparison. 

\begin{figure}
    \centering
    \includegraphics[width=\columnwidth]{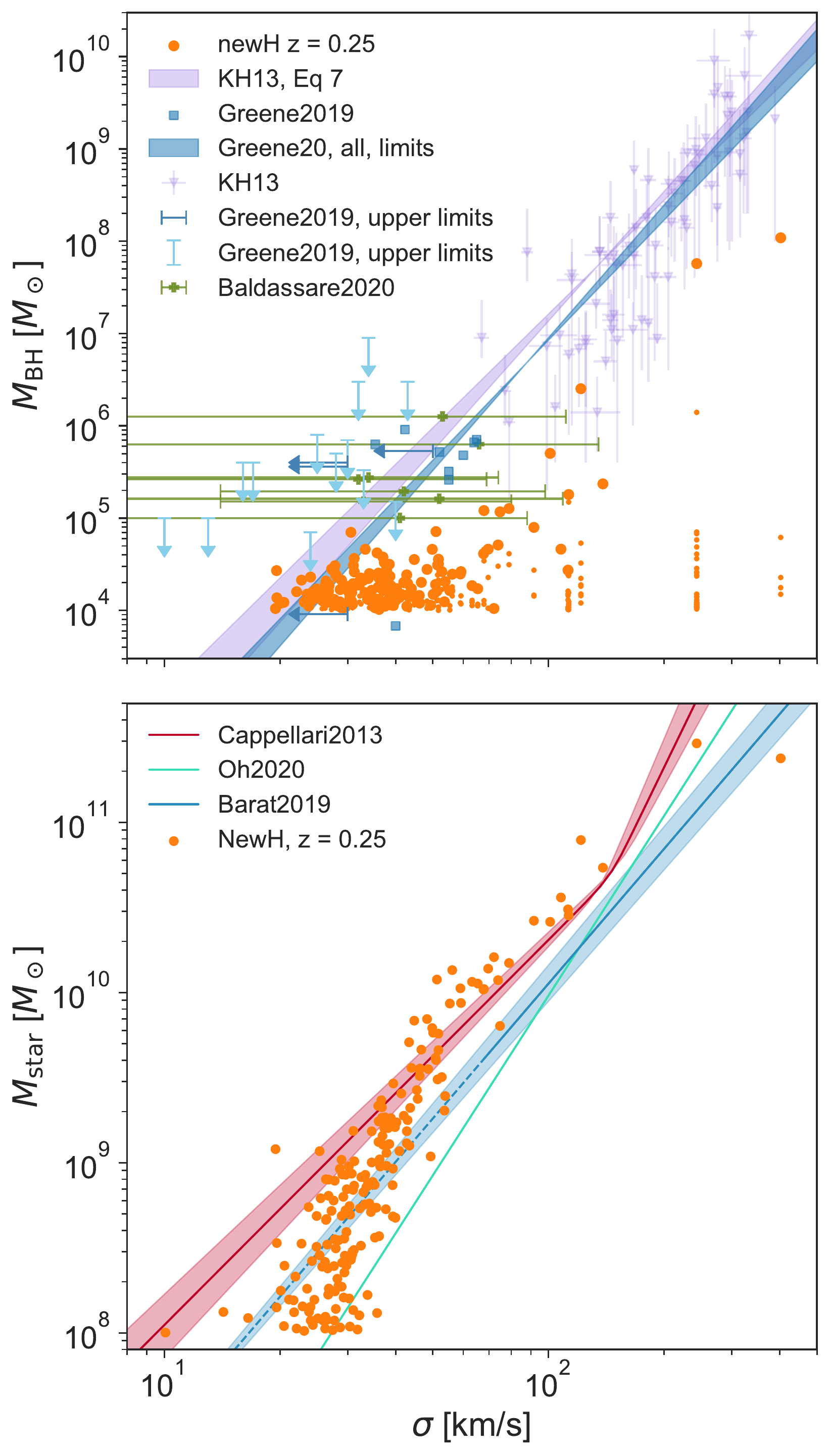}
    \caption{[Top] BH mass versus stellar velocity dispersion $\sigma$, for all galaxies with a stellar mass above $10^8 \msun$ at $z=0.25$. Observations shown are from \citet{Greene2020}(Greene2019, blue markers), \citet{Baldassare2020} (Baldassare2020, green crosses) and \citet{Kormendy2013} (KH13, purple markers).  [Bottom] Faber-Jackson relation for redshift $z=0.15$. Shown for comparison are observational fits from from \citet{cappellarietal13,baratetal19} and \citet{ohetal20}.}
    \label{fig:sigma_relations}
\end{figure}

Alternatively BH growth in galaxies in \nh~might be artificially suppressed due to an AGN feedback efficiency that could have been set too high, as the BH feedback efficiency controls the normalisation of the correlations between BH and their host galaxy properties \cite{Dubois2012}. Evidence for this hypothesis comes from both the $M_{\rm BH}-M_{\rm star}$ relation (top panel, Fig. \ref{fig:MBH_Mstar}) and the $M_{\rm BH}- \sigma$ relation (top panel, Fig \ref{fig:sigma_relations}) where BHs in \nh~fall consistently below observational values and fits. The bottom panel of Fig \ref{fig:sigma_relations} confirms that this is not due to the fact that stellar velocity dispersions $\sigma$ are systematically over-estimated in \nh, as galaxies at this redshift fall within observational constraints. While BHs seem to fall on the observed $M_{\rm BH}$-$\sigma$ relation at the low-mass end, we caution that the mass of any BH that has not grown by at least an order of magnitude is likely to still be dominated by the numerical seed mass. 
  
Another possible explanation for the lack of BH growth in \nh~ is due to the position of BHs in galaxies. It is suspicious that the slope of the $M_{\rm BH,acc} - M_{\rm star}$ relation is very similar for main BH in dwarf galaxies and secondary BH in massive galaxies, which could suggest that BH in dwarf galaxies fail to grow as they are insufficiently close to the centre of their host galaxy. This effect in \nh~ might be enhanced due to the similarity of the BH seed mass, $M_{\rm BH,0}=10^4 \rm \ M_\odot$, and the stellar mass resolution of $1.3 \times 10^ 4 \rm \ M_\odot$, which leads to artificial scattering of BH trajectories by interactions with the stellar particles although this two-body effect is somewhat mitigated by the multi-grid gravity solver. We explore the spatial distribution of BH, and its impact on BH growth, further in Sec. \ref{sec:orbits}.
  
Whether the break in the $M_{\rm BH}-M_{\rm star}$ relation is real remains an open question. With its lower seed mass of $10^4 \rm \ M_\odot$, an order of magnitude lower than previous simulations of dwarf galaxies \citep[see e.g.][ who used a seed mass of $10^5 \rm \ M_\odot$]{Koudmani2021,Sharma2022} and two orders of magnitude lower than the typical seed masses of $10^6 \rm \ M_\odot$ of large-scale cosmological simulations \citep[see][for a comparison of different simulations]{Haidar2022}, \nh~probes the co-evolution between IMBHs and dwarf galaxies for a larger range of galaxy masses. Using a seed mass of $10^6 \rm \ M_\odot$, \citet{Koudmani2021} find tentative evidence for a flattening of the $M_{\rm BH} - M_{\rm star}$ relation, while \citet{Sharma2022} do not, but instead discuss that this could be an artifact of their high seed mass, which artificially boosts accretion onto overmassive BHs in galaxies.

In the rest of the paper, we will explore what drives the (lack of) mass growth of BHs in dwarf galaxies, quantify whether the observable population of IMBHs in \nh~reproduces observed AGN in dwarf galaxies, and how the population of observable IMBHs in dwarf galaxies compares to the full sample.

\subsection{Occupation fractions} 
\label{sec:occupation}

\begin{figure}[h]
    \centering
    \includegraphics[width=\columnwidth]{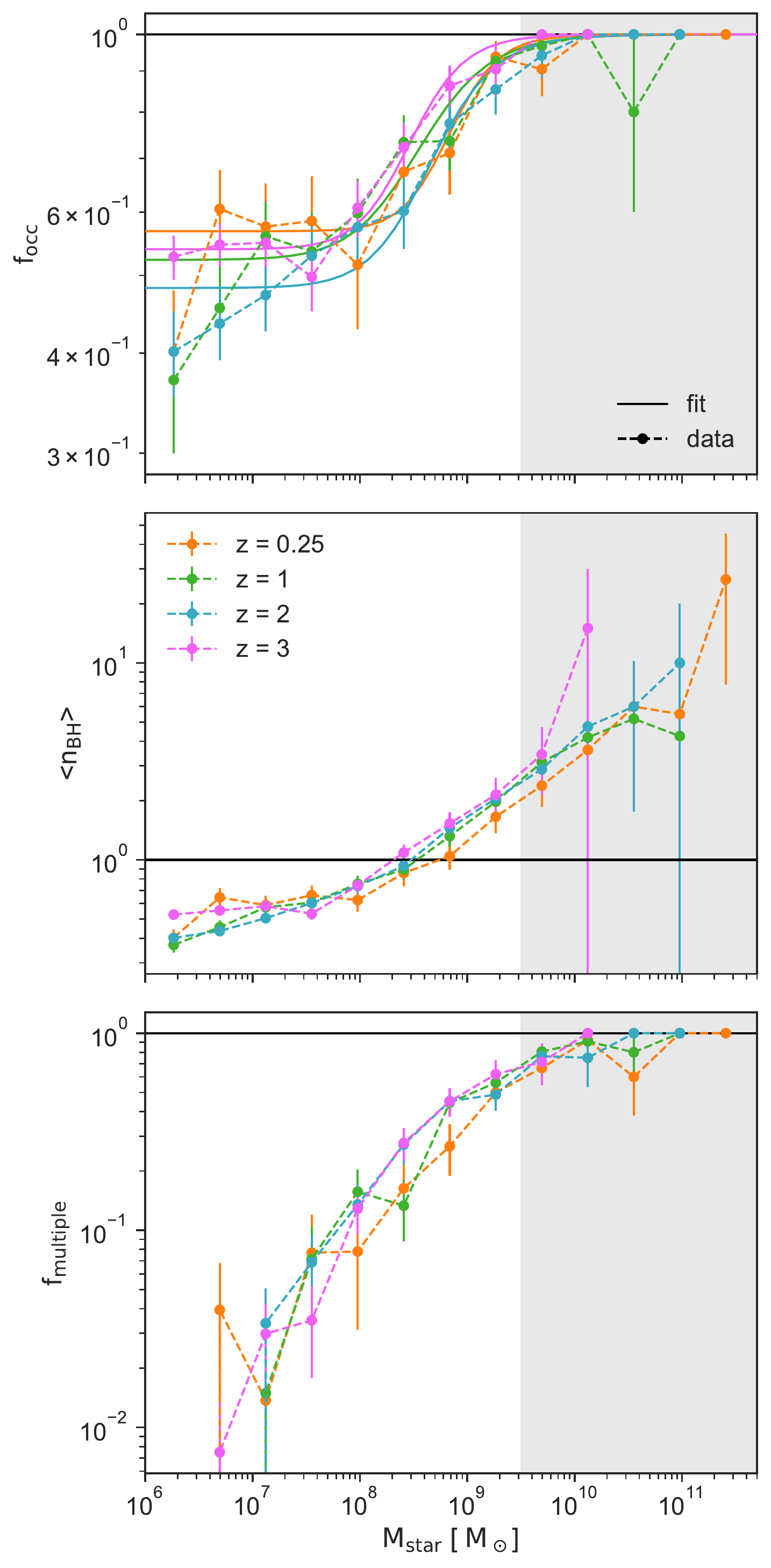}
    \caption{Average number of BHs per galaxy [top], fraction of occupied galaxies [middle] and fraction of galaxies with multiple BHs [bottom] as a function of galaxy mass $\rm M_{\rm star}$. A white background denotes dwarf galaxies. Solid lines on the top panel denote a fit of Eq. \ref{eq:fit_occupation_fraction}, with free parameters at each redshift listed in Table \ref{tab:fit_occupation_fraction}. Dashed lines connect data points. Error bars show Poisson errors.}
    \label{fig:occupation_fraction_galaxies}
\end{figure}

One key question to understand the population of IMBHs in dwarf galaxies is how many dwarf galaxies contain a main BH. There are several ways to measure occupation fractions, as illustrated by the different panels in Fig. \ref{fig:occupation_fraction_galaxies} (as a function of galaxy mass). The top panel shows the fraction of galaxies (halos) that contain at least one BH, the second panel takes BH multiplicity into account and shows the average number of BH per galaxy (halo), while the third panel denotes the fraction of galaxies that contain two or more BHs. In this section, we discuss occupation fractions of galaxies and halos which measure the presence of BHs in a given galaxy or halo, regardless of their luminosity. For the fraction of active BHs (AGN fractions), see Section \ref{sec:AGN_frac}.

As already seen in the $M_{\rm BH}-M_{\rm star}$ relation in Fig. \ref{fig:MBH_Mstar}, BHs are formed in galaxies with stellar masses as low as $M_{\rm star} = 10^6 \rm \ M_\odot$. As can be seen in Fig. \ref{fig:occupation_fraction_galaxies} occupation fractions for galaxies and halos fall with decreasing galaxy (halo) mass but remain significant across the dwarf galaxies, with even the lowest mass galaxies having a minimum occupation fraction of $\rm f_{\rm occ} > 0.37$. There is little redshift evolution for all BHs across the mass range, with 57\% of dwarf galaxies containing at least one BH at $z=3$, in comparison to 54 \% at $z=0.25$. Multiple occupation of galaxies is fairly common for massive dwarfs (with more than 26 percent of dwarf galaxies with $M_{\rm star} > 10^{8} \rm \ M_\odot$ hosting more than one BH at $z=0.25$), but becomes increasingly uncommon at lower galaxy stellar mass. The lack of growth of IMBHs in dwarf galaxies is therefore not due to low number statistics. \nh~contains 376 BHs distributed across 308 dwarf galaxies, out of a total of 552 dwarf galaxies at $z=0.25$.

In comparison to \citet{Sharma2022}, who use the Romulus25 simulation to study the population of BHs in dwarf galaxies,  we also report a sharp decline in the occupation fraction with galaxy stellar mass at low redshift. At higher redshift, however, occupation fractions in \nh~are considerably flatter that in Romulus25. This could be due to the lower stellar mass resolution in \nh~ ($1.3 \times 10^4 \rm \ M_\odot$) in comparison to Romulus25 ($2.1 \times 10^5 \rm \ M_\odot$), and the corresponding lower seed mass ($10^4 \rm \ M_\odot$ in \nh~in comparison to $10^6 \rm \ M_\odot$ for Romulus25), which allows the coevolution of a given galaxy and its BH to be followed from earlier on in its evolution. As the intrinsic occupation fraction cannot be observed, we defer a comparison to observations to the fraction of active BHs in Section \ref{sec:AGN_frac}.

For ease of comparison to other datasets, the fraction of occupied galaxies has been fit with a function of the form
\begin{equation}
    f(M)=1-\frac{\alpha}{1+\left(\frac{\log{(M)}}{M'}\right)^\epsilon}
    \label{eq:fit_occupation_fraction}
\end{equation}
where $M$ is either the galaxy stellar mass $M_{\rm star}$ or the halo mass $M_{\rm halo}$, and the free parameters of the fit $\alpha$, $\epsilon$ and $M'$ are listed in Table \ref{tab:fit_occupation_fraction} for each redshift and both galaxy samples.

\begin{table}
    \centering
    \begin{tabular}{ l ccc}
        \hline
        \textbf{z} &  $\alpha$ & $M'$ & $\epsilon$  \\
        \hline
        $z=0.25$ & 0.48 & 8.84 & 21.56 \\$z=1$ & 0.64 & 8.2 & 9.55 \\$z=2$ & 0.6 & 8.6 & 11.89 \\$z=3$ & 0.47 & 8.54 & 28.45 \\
        \hline
    \end{tabular}
    \caption{Fitting parameters for $\rm f_{\rm occ}$ from Eq. \ref{eq:fit_occupation_fraction}. Fits are shown in the top panel of Fig. \ref{fig:occupation_fraction_galaxies} as solid lines.}
    \label{tab:fit_occupation_fraction}
\end{table}

\subsection{Black hole mass functions}
\label{sec:mass_function}

 \begin{figure}
    \centering
    \includegraphics[width=\columnwidth]{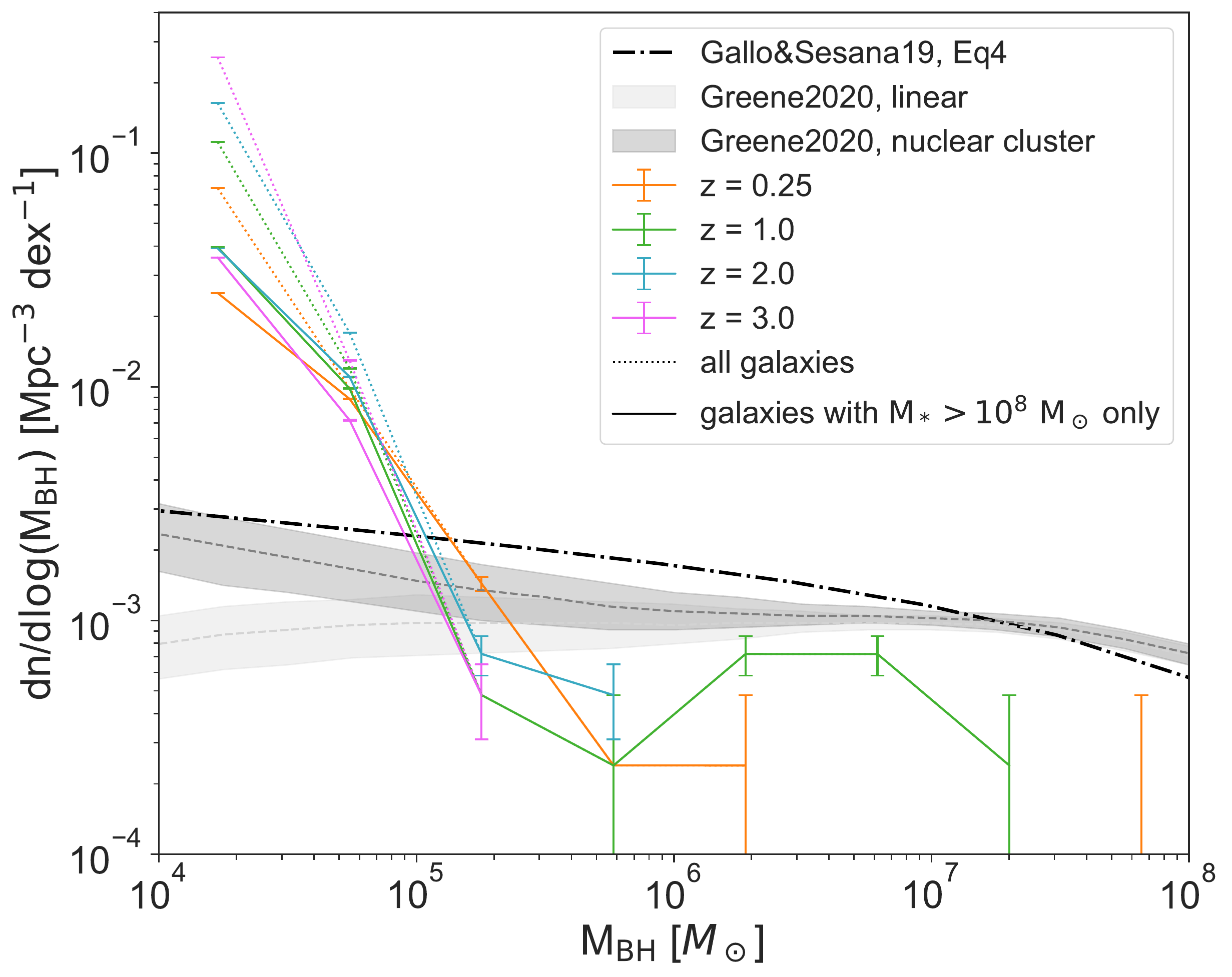}
    \caption{Evolution of the mass function of main BHs with redshift. Shown in comparison are fits based on observations from \citet{Gallo2019} (black dotted) and \citet{Greene2020}(grey shaded). The two models for \citet{Greene2020} are derived using a linear occupation fraction (light grey) and a nuclear cluster occupation fraction (dark grey) respectively. BH mass functions are annotated with Poisson error bars. Solid lines include main BHs of galaxies with a stellar mass above $10^8 \msun$, while dotted lines include all main BHs.}
    \label{fig:number_density_z}
\end{figure}

One consequence of the efficient seeding and low BH growth in dwarf galaxies is that there is a large number of BHs at or near the seed mass. As a result, the  BH mass function shown in Fig. \ref{fig:number_density_z} steepens strongly at low BH masses at all redshifts,  in comparison to predictions based on observed galaxy stellar mass function (grey and black lines from \citealp{Gallo2019} and \citealp{Greene2020} respectively). Even when restricting our BH sample to more readily observable objects, by taking only main BHs of galaxies with stellar masses above $10^8 \msun$ into account (solid line), our BH mass functions remain very high for BHs below $10^5 \msun$. This is due to a combination of an efficient seeding algorithm, which creates high occupation fractions of BHs in low mass galaxies, and lack of sustained growth of the BH. 

Conversely, \nh~under-predicts the number of more massive BHs, i.e. falls below observational limits for $M_{\rm BH} > 10^5 \  \msun$, suggesting that despite ample seeding, BHs struggle to grow beyond their seed masses in the environment probed here. We note that due to our comparatively small simulation volume combined with the fact that \nh~probes an average volume rather than an overdense one, number statistics for massive galaxies and their high mass BH are poor, and errorbars therefore comparatively large. We observe no clear evolution of the BH mass function with redshift for the mass bins probed at a given redshift, with the only evolution coming from BHs growing to a more massive regime over time.

\subsection{Black holes in dwarf galaxies as AGN}
\label{sec:bolometric_luminosity}

\begin{figure*}
    \centering
    \includegraphics[width=\textwidth]{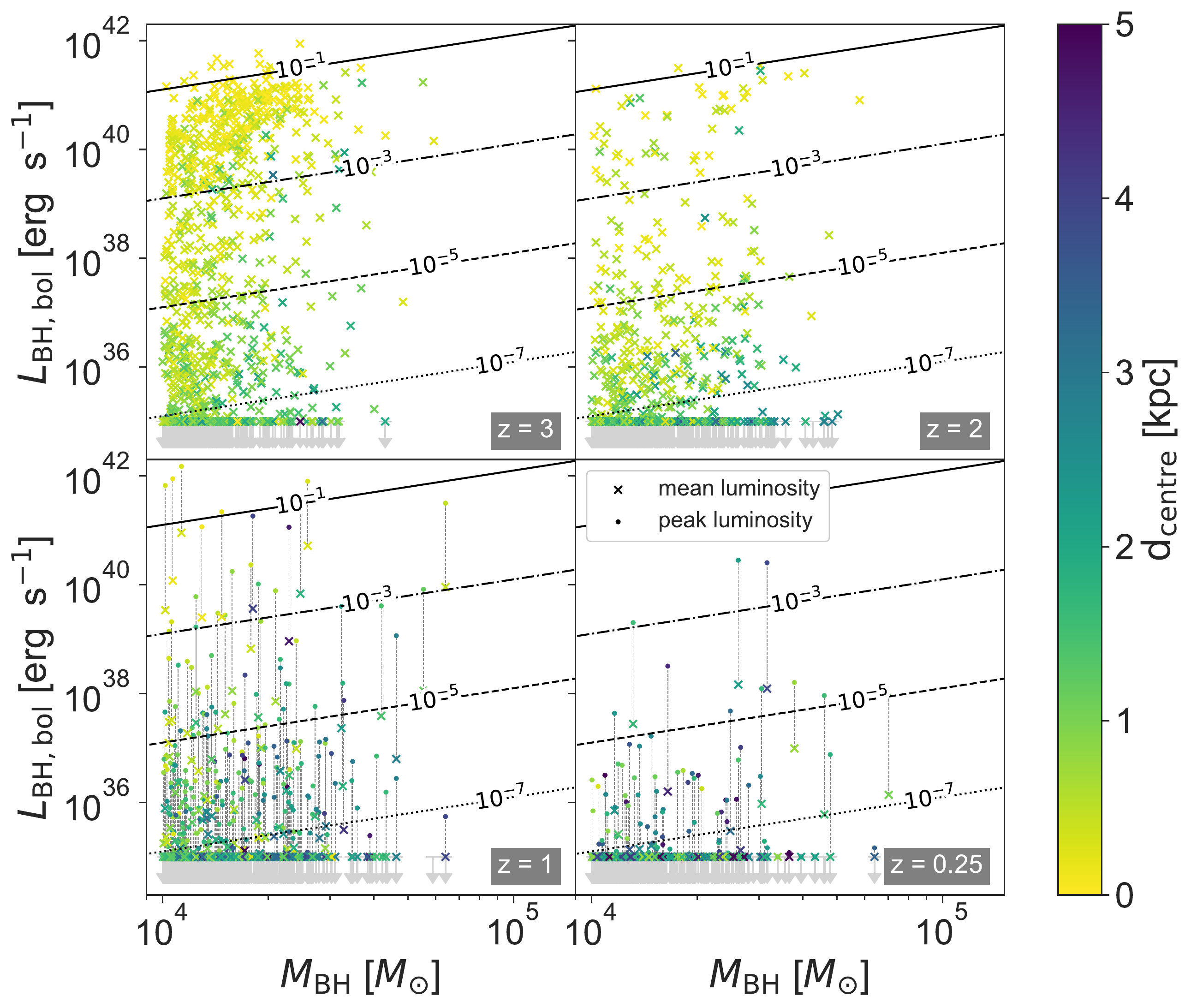}
    \caption{BH bolometric luminosity versus BH mass for all main  BHs in dwarf galaxies,  i.e., with stellar mass below $M_{\rm dwarf} = 3 \times 10^9 \rm \ M_\odot$. Points are coloured by the distance between the BH and the centre of the host galaxy. Mean luminosities over $\pm \Delta t = 100 \rm \ Myr$ are shown for all redshifts (crosses), with BHs with mean $L_{\rm bol} < 10^{35} \rm \ erg \ s^{-1}$ plotted at $10^{35} \rm \ erg \ s^{-1}$ as upper limits. For $z\leq 1$, peak luminosities over the same $\Delta t$ are also shown (dots). Constant Eddington fractions  $f_{\rm Edd}$ are shown as black lines.} 
    \label{fig:fedd}
\end{figure*}

As can be seen in Fig. \ref{fig:fedd}, the ability of BHs to accrete effectively in dwarf galaxies, and the associated luminosity of AGN in dwarf galaxies, evolves strongly with redshift. In this section we differentiate between the instantaneous bolometric luminosity of a BH, $L_{\rm BH,bol}$, measured at the target redshift, a mean bolometric luminosity of a BH, $L_{\rm BH,bol,mean}$, which is calculated by taking the mean of $\log(L_{\rm BH,bol})$ for all values within $\Delta t \pm 100 \rm \ Myr $ around the target redshift, and a peak bolometric luminosity, $L_{\rm BH,bol,peak}$ which is defined to be the maximum luminosity of a BH within $\Delta t \pm 100 \rm \ Myr $ around the target redshiftWe used the BH spin based radiative efficiencies to compute the BH luminosities, as detailed in \citet{Dubois2021}. As we showed in the companion paper on BH spin for \nh~BHs \citep{Beckmann2022d}, spin-based radiative efficiencies do modulate the bolometric luminosity for individual BHs by up to a factor of $\sim 3$, but have on average little impact on the distribution of luminosities for the whole population. This increased feedback efficiency, and consequent reduced mass growth of BHs, could explain why all main BHs in massive galaxies are found at the lower end of observations.

At high redshift, there is a wide range of accretion efficiencies across the sample of BHs. The most efficiently accreting BH growth with time-average Eddington ratios $f_{\rm Edd,mean} = \bar{L}_{\rm BH,bol,mean}/L_{\rm Edd}>0.1$ but such BH are rare as the majority of the sample grow so inefficiently that their mass growth is negligible: 76.6 percent have a $f_{\rm Edd}\leq3 \times 10^{-3}$, i.e. a mass growth timescale greater than the Hubble time and a luminosity far too faint to be detectable in AGN surveys. However, even among low accretors, brief accretion spikes with significantly higher luminosities are common: 23.6 percent of BHs at $z=3$ have peak luminosities of $f_{\rm edd,peak}>0.1$, compared to only 1.3 percent of the sample whose $f_{\rm edd,mean}$ is that high. If such accretion spikes leave a longer-lasting signal, they might enhance the observability of high-redshift BH. 

With decreasing redshift, BH activity decreases markedly. At $z=0.25$, no BH in a dwarf galaxy has a mean Eddington fraction larger than $f_{\rm Edd} = 10^{-4}$ which translates an effective minimum Salpeter time of 450 Gyr: by $z=0.25$, growth for BHs in dwarf galaxies has effectively stopped in \nh. It is well known that BH growth slows down over time \citep[][]{Dubois2012,Volonteri2016,Habouzit2022}. However, there is growing evidence that BH growth in simulated dwarf galaxies slows down even faster than the slowdown in stellar mass growth, as similar trends to those in \nh~ are also found in other simulations of IMBH in dwarf galaxies, \citep[][]{Barai2019,Koudmani2021,Sharma2022}, who all concluded that AGN activity in dwarf galaxies decreases strongly with redshift. A question that remains to be addressed is if, statistically, BH growth in simulated dwarf galaxies slows down faster than for more massive galaxies. 

As can be seen by the colour-coding of datapoints in Fig. \ref{fig:fedd}, the location of BHs plays a role in their decreased activity at low redshift: efficiently growing BHs are on average very close to the centre of their host galaxies, while the accretion efficiency of those further out drops markedly. We discuss the impact of BH location within their host galaxy further in Sec. \ref{sec:orbits}. 

\begin{figure*}
    \centering
    \includegraphics[width=\textwidth]{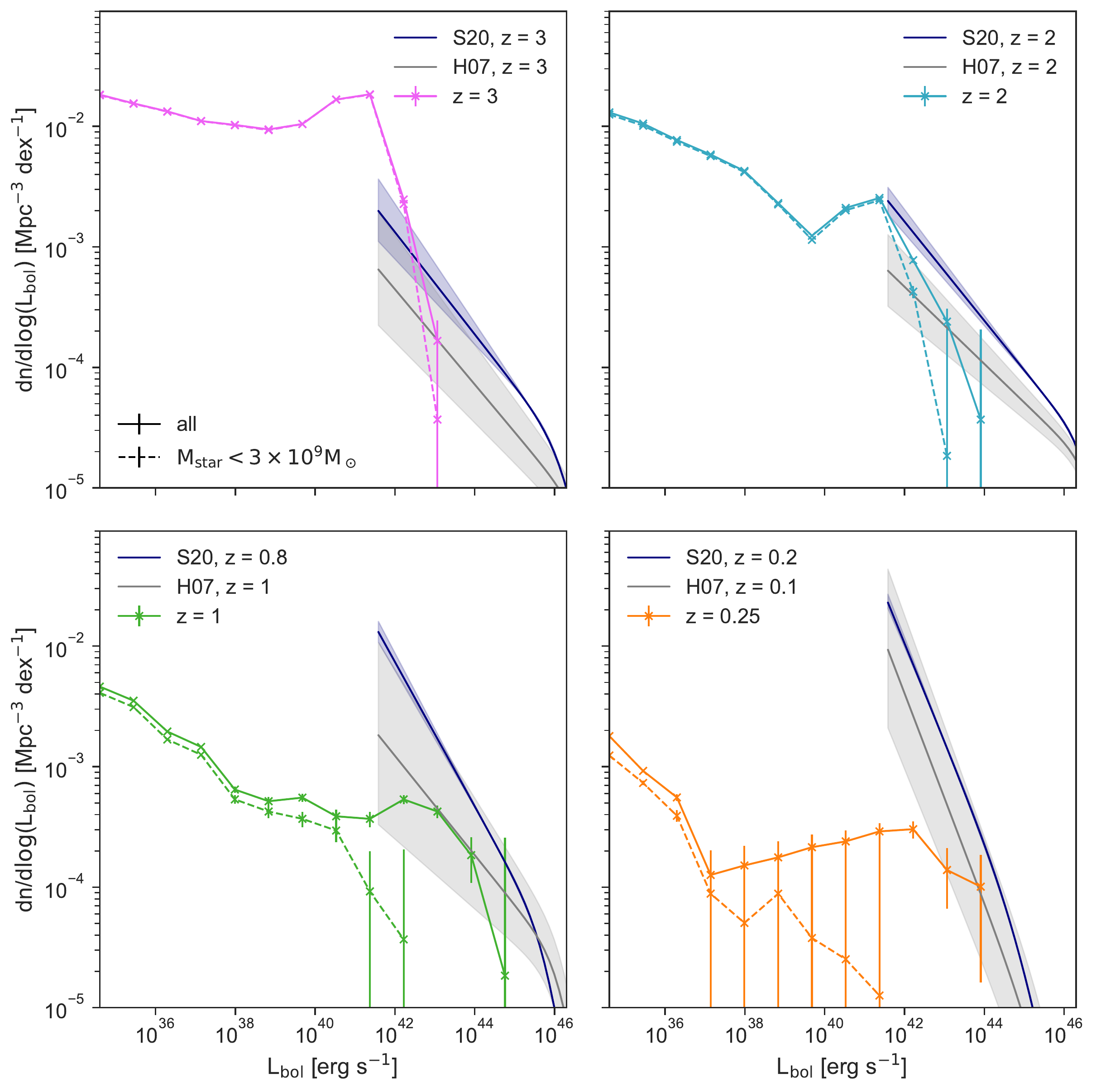}
    \caption{Bolometric BH luminosity functions with Poisson error bars for the full sample of galaxies (solid) and dwarf galaxies (dashed). Luminosities shown are derived from a stacked sample within $\pm 100$ Myr of the target redshift. Shaded regions show fits to observational luminosity functions by \citet{Hopkins2007} (H07) and \citet{Shen2020} (S20) for comparison.}
    \label{fig:luminosity}
\end{figure*}

Despite their low average growth rates, BHs continue to see brief bursts of activity even at low redshift, as can be seen by the peak luminosities plotted in the lower two panels of Fig. \ref{fig:fedd}, which can be more than an order of magnitude higher than the mean luminosity. As a result, the bolometric luminosity functions in Fig. \ref{fig:luminosity} do not drop to zero even at low redshift.  At each redshift in Fig. \ref{fig:luminosity}, luminosities plotted are derived from a stacked sample of $L_{\rm BH,bol}$ within $\Delta t = \pm 100$ Myr of the target redshift, to account for the variation in AGN luminosity. Results do not depend sensitively on the choice of $\Delta t$. Luminosity functions are shown for both the sample of  all galaxies in \nh, and for dwarf galaxies only. Due to the mass evolution of galaxies over time,  the two samples are indistinguishable at high redshift ($z \geq 2$) except at the very bright end, but the decrease in AGN activity for BHs in dwarf galaxies can be clearly seen at $z\leq 1$.

At high redshift, the luminosity function shows a clear peak around $10^{42} \rm \, erg \, s^{-1}$, which is approximately equal to the Eddington luminosity for our seed BHs. The peak in the luminosity function at this luminosity shows that many BHs at $z \geq 2$ are accreting as efficiently as permitted by our accretion algorithm (see Sec. \ref{sec:simulation} for details) One intriguing possibility is that BHs in \nh~are undermassive at low redshift because accretion is capped at the Eddington limit at high redshift. Even brief super-Eddington episodes could give BHs an early mass boost  \citep{Regan2018} that could influence their evolution later, although \citet{Massonneau2022} have shown that super-Eddington accretion actually reduced the overall growth of BHs in massive compact galaxies.

By $z<1$, the peak in the luminosity function around $10^{41}-10^{42} \ \rm erg s^{-1}$ has disappeared. At this redshift, the luminosity function at $L_{\rm bol} < 10^{41} \rm\, erg \, s^{-1}$ is still almost entirely due to BHs in dwarf galaxies, while BHs in massive galaxies dominate at higher luminosity. By $z=0.25$, BHs in massive galaxies dominate as far down the luminosity function as $10^{37} \rm \, erg\, s^{-1}$, with only a small contribution from BHs in dwarf galaxies in this luminosity range due to brief peaks of AGN activity (see Fig. \ref{fig:fedd}). At all redshifts, the bright end ($\rm L_{\rm bol} > 10^{41} \rm \, erg \, s^{-1}$) of the luminosity function is generally lower than the values by \citet{Shen2020} but in good agreement with observations by \citet{Hopkins2007}, with the exception of $z=0.25$ where \nh~under-predicts the luminosity function for $L < 10^{43} \rm \, erg \, s^{-1}$. This most likely simply reflects the fact that there are no overmassive BHs in \nh~and many galaxies host undermassive BHs (see Fig. \ref{fig:MBH_Mstar}). We note that given that \nh~ is a zoom simulation, luminosity functions should be treated with caution as we do not have a statistically significant sample of massive galaxies, and their SMBH. As the low-luminosity end of the luminosity function could be populated by both highly accreting IMBH and inefficiently accreting SMBH, the absence of SMBH in the sample will be felt across the whole luminosity function.

\subsection{AGN fractions}
\label{sec:AGN_frac}

\begin{figure*}
    \centering
    \includegraphics[width=\textwidth]{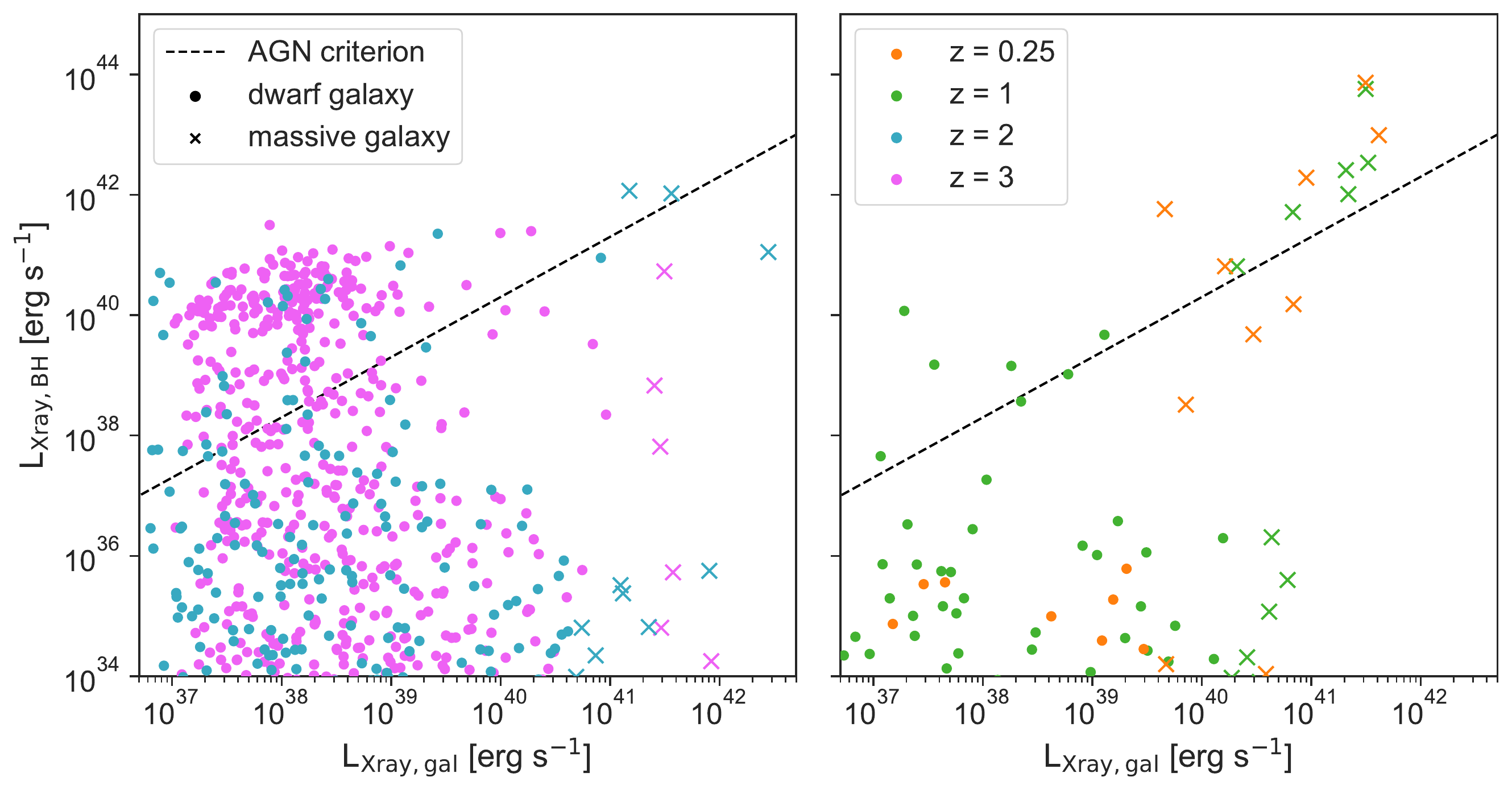}
    \caption{Distribution of X-ray luminosity of BHs versus that of their host galaxies using the mean BH luminosity (left, $z=0.25,1,2 \& 3$) and peak BH luminosity (right, $z=0.25 \& 3$ only) within a $\Delta t \pm 100\rm \  Myr$ window.  Data-points above the dashed line are detectable as AGN, as they have $L_{\rm Xray, BH} > 2 L_{\rm Xray,gal}$, if the combined luminosity of BH and galaxy exceeds the luminosity cut, i.e. if also  $L_{\rm Xray, BH} + L_{\rm Xray, gal} > L_{\rm cut}$ for a given $L_{\rm cut}$.}
    \label{fig:Lxray}
\end{figure*}

To understand how the population of BHs in \nh~compares to observed populations of IMBHs in dwarf galaxies, we look at the fraction of BHs that would be observable as AGN in the X-ray. 

For the BH X-ray luminosity, we calculate the $0.5-10 \rm \ keV$ luminosity using  the bolometric correction from \citet{Shen2020}. For each galaxy, the BH X-ray luminosity $L_{\rm Xray,BH}$ is the sum of the instantaneous X-ray luminosities of all BHs associated with that galaxy, and is, hence, the sum of the galaxy's main BH as well as its secondary BHs. $L_{\rm Xray,BH}$ is therefore the BH luminosity that would be observed using a telescope with insufficient angular resolution to separate individual BHs in the galaxy. Typically, $L_{\rm Xray,BH}$ is dominated by the main BH.  This is partially because not every galaxy even has a secondary BH (see the discussion on BH multiplicity in Section \ref{sec:occupation}) and because main BHs are much more active: main BH are 2 ($z\geq1$) to 4 ($z=0.25$) times more likely to appear on Fig. \ref{fig:Lxray}, i.e. to have an instantaneous X-ray luminosity in excess of the solar luminosity $10^{34} \rm \ erg \, s^{-1}$ than secondary BHs. As a result, only 5 percent ($z=0.3$) to 11 percent ($z=0.25$) of galaxies have secondary BHs that contribute more than 20 percent to $L_{\rm Xray, BH}$.

To assess the detectability of our AGN, we also compute the X-ray luminosity of the host galaxy from both X-ray binaries (XRBs) and hot gas. For XRBs, we compute the total X-ray luminosity  $L_{\rm XRB}$ by adding the contributions from soft ($0.5-2 \rm \ keV$) and hard ($2-10 \rm \ keV$) X-rays according to the redshift dependent relation from \citet{Lehmer2016}. For X-ray emission from hot gas $L_{\rm gas}$ in the host galaxy, we compute the emission in the soft X-ray band using the relation from \citet{Mineo2012}, and extrapolate to the hard X-ray band assuming a photon index of $\Lambda = 3$ following \citet{Mezcua2018}. The total X-ray luminosity of the host galaxy is $L_{\rm Xray,gal}= L_{\rm XRB} + L_{\rm gas}$. A system is considered observable for a given luminosity cut $L_{\rm cut} $ if $L_{\rm Xray, BH} + L_{\rm Xray, gal} > L_{\rm cut}$, and detectable as an AGN if also $L_{\rm Xray, BH} > 2 L_{\rm Xray,gal}$ following \citet{Birchall2020}. The resulting distribution of $L_{\rm Xray,gal}$ versus $L_{\rm Xray, BH}$ can be seen in Fig. \ref{fig:Lxray} for different redshifts.

The number of BHs that meet the \citet{Birchall2020} criterion decreases strongly with redshift: while there are plenty of BHs identifiable as AGN at $z\geq 2$, there are no BHs in dwarf galaxies at $z=0.25$ that would be recognisable as AGN using the \citet{Birchall2020} requirement (dashed line), no matter the luminosity cut. This would remain true in even the most optimistic case if we considered the peak BH luminosity within a $\Delta t \pm 100 \rm \ Myr$ window around the target redshift (not shown on the plot).

\begin{figure}
    \centering
    \includegraphics[width=\columnwidth]{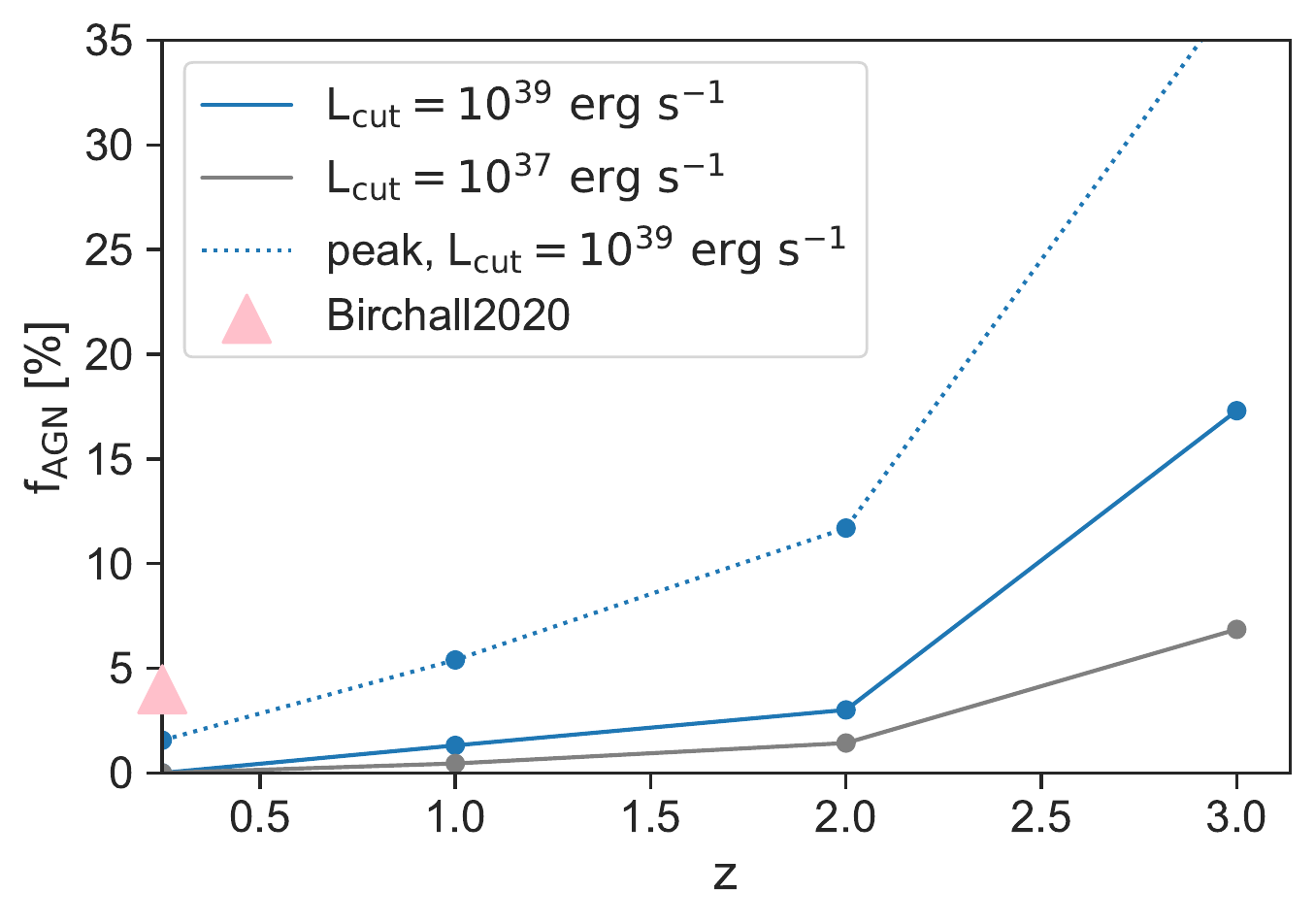}
    \caption{Fraction of observable AGN as a function of redshift for all dwarf galaxies with a total X-ray luminosity above $10^{39} \rm \ erg\, s^{-1}$ (blue) and  $10^{37} \rm \ erg\, s^{-1}$ (grey)  
    Solid lines use the instantaneous $L_{\rm xray,BH}$ at $z=0.25$, while dotted lines use the peak $L_{\rm xray,BH}$ within $\Delta t \pm 100 \rm \ Myr$}. Coloured markers show observational data measured at $z<0.25$  by  \citet{Birchall2020}, for luminosity cuts of $>10^{39} \rm \ erg\, s^{-1}$.
    \label{fig:fAGN_z}
\end{figure}

As a result, the fraction of AGN in dwarf galaxies drops strongly with decreasing redshift, as can be seen in Fig. \ref{fig:fAGN_z}, from $f_{\rm AGN} \sim 17 $ percent at $z=3$ to $f_{\rm AGN} =0 $ percent at $z=0.25$ using the instantaneous X-ray BH luminosity $L_{\rm BH,xray}$ and a luminosity cut of $L_{\rm cut} > 10^{39} \rm \ erg\, s^{-1}$. This is in line with previous simulation results by \citet{Barai2019} and \citet{Koudmani2021}, who both showed that AGN activity in dwarf galaxies is significantly higher at high redshift. Like in \nh, there are no remaining active BHs in dwarf galaxies in the FABLE simulations at $z\leq 0.25$ \citep{Koudmani2021}. This disagrees with observations from \citet{Birchall2020}, who report an AGN fraction of 4 \% at $z<0.25$. We note that the observed AGN fractions in dwarf galaxies in literature vary strongly depending on how an AGN is defined and how the X-ray luminosity of the host galaxy is accounted for. As a result, some X-ray surveys find values as high as 30 percent  \citep{Zhang2009,She2017}. Here we restrict the comparison to \citet{Birchall2020} as we have designed our AGN observability criteria to reflect theirs.

AGN fractions depend strongly on the luminosity cut, as can be seen by comparing the $10^{39} \rm \ erg s^{-1}$ to the even more optimistic $10^{37} \rm \ erg s^{-1}$, which shows lower $f_{\rm AGN}$ at all redshifts probed here. This is due to the fact that with increasing luminosity cut, the number of AGN in the sample increases but the number of galaxies increases even faster. A direct comparison to observations of this redshift evolution is difficult as observing AGN with luminosity cuts as low as $10^{39} \rm \ erg\, s^{-1}$, let alone the lower cut of $10^{37} \rm \ erg\, s^{-1}$, is currently only possible at very low redshift. Instead, observations of the redshift evolution of $f_{\rm AGN}$ such as those presented in \citet{Mezcua2018} use luminosity cuts of the order of $10^{40.5} \rm \ erg\, s^{-1}$, which no BH in dwarf galaxies achieve in \nh, even at $z=3$. At such higher luminosity cuts, observations show a predominantly flat evolution of the AGN fraction with decreasing redshift at $z<1$ \citep{Mezcua2018,Birchall2020}, which suggests that current simulations over-quench AGN. This hypothesis is further supported by the fact that there are no bright AGN at all in \nh. This conclusion assumes that AGN are observed at their instantaneous luminosity at $z=0.25$. If we make the generous assumption that AGN activity leaves an observable signal that persists for longer than the AGN outburst itself, and that we can therefore observe AGN at their peak activity within a $\Delta t = 100 \rm \ Myr$ window (similar to the analysis in Fig. \ref{fig:luminosity}) then two potential AGN do appear in the sample at $z=0.25$. This raises the observable AGN fraction to $f_{\rm AGN} = 1.5 \%$. Despite this being a generous assumption, the resulting AGN fraction still lies below the observed fraction of 4 \% by \citet{Birchall2020}. We also caution that this  assumes that the observable AGN signal does not fade at all for up 100 Myr after the outburst has ended, and therefore represents an upper limit.

Overall, we conclude that mass growth for BH with seed masses of $10^4 \rm \ M_\odot$ in \nh~ is stifled below observed limits, particularly at low redshift.  The time window for efficient BH growth is too short for BHs to compensate for the lack of earlier growth at the seed mass.
This could suggests that BHs seeds in dwarf galaxies are formed as massive seed BHs ($>10^4\, \rm \ M_\odot$), or that current SN prescriptions are too strong to allow for the observed BH growth. Other potential numerical effects, besides SN prescriptions, that might lead to the observed over-quenching, is an over-estimated AGN feedback efficiency. However, it is not as simple as simply decreasing the SN feedback strength, as doing so will over-predict the stellar-to-halo mass relation, which for \nh~is already at the upper end of empirical constraints. Instead, it might be more a question of where SN energy is deposited, how turbulence is injected in the interstellar medium and how SN and possibly AGN feedback can continue to regulate the galaxy mass content without over-quenching IMBHs in dwarf galaxies~\citep[see][for a detailed investigation of the impact of SN feedback strength on IMBH growth in dwarf galaxies]{Koudmani2022}. Alternatively, the lack of BH growth could be a sign of other relevant missing physics, such as the suppression of cooling from cosmic rays which have been shown to suppress the star formation rate (SFR), and resulting SN injection rate, by a factor of 2-3 \citep{Dashyan2020}.

\subsection{Distribution of black holes in dwarf galaxies}
\label{sec:orbits}

\begin{figure}
    \centering
    \includegraphics[width=\columnwidth]{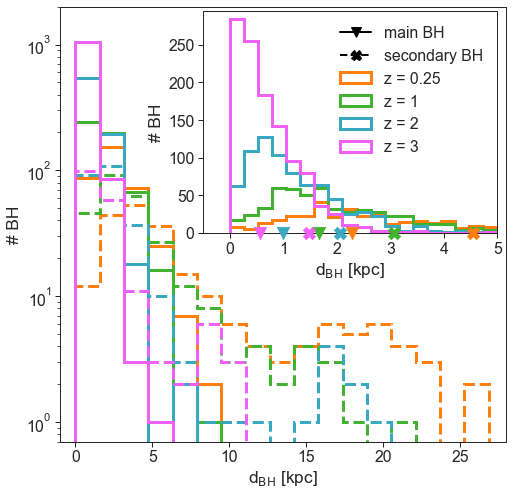}
    \caption{Distribution of the distance between BHs and the centre of their host galaxies, $d_{\rm BH}$ for main (solid) and secondary (dashed) BHs.  The in-set pan shows a zoom in on the range $d_{\rm BH} < 5 \rm \ kpc $ and the median distance of main (secondary) BHs as triangles (crosses) at each redshift.}
    \label{fig:BH_pos}
\end{figure}

To grow efficiently, BHs must be able to tap into an abundant local supply of cold gas. The lack of growth for BHs in \nh~could therefore be due to one of two reasons: either dwarf galaxies in \nh~are cold gas poor, or BHs are not located where the cold gas is. As can be seen by the colour-coding of datapoints in Fig. \ref{fig:fedd}, the location of BHs certainly plays a role in their decreased activity: efficiently growing BHs are on average very close to the centre of their host galaxies, while the ability to accrete of those further out drops markedly. The distribution of BHs from the centre of their host galaxy evolves with redshift, as can be seen qualitatively in Fig. \ref{fig:fedd} and more quantitatively in Fig. \ref{fig:BH_pos}. The sample of BHs plotted in Fig. \ref{fig:BH_pos} separately analyses the distribution  of main and secondary BHs, unlike Fig. \ref{fig:Lxray} which only shows main BHs. At all redshifts, main BHs dominate the sample, with 86\% of BHs at $z=3$ classified as `main' which decreases to 62\% at $z=0.25$.

In general, main BHs are closer to the centre of their host galaxy than secondary BHs, as can be seen in Fig. \ref{fig:BH_pos}. We note that main BHs being closer to the centre of their host galaxy is not by design as the `main' BH of a galaxy is not selected to be the one  located closest to the centre of the galaxy. Instead, it is defined to be the most massive BH that meets the criterion of being identified with a given galaxy (see Sec. \ref{sec:catalogue} for more details).

Both main BHs and secondary BHs become less centrally located over time. The median distance of BHs to the centre of their galaxy increases from $0.68 \rm \ kpc$ at $z=3$ to $2.96 \rm \ kpc$ at $z=0.25$ for main BHs, and from $1.48 \rm \ kpc$ at $z=3$ to $4.55 \rm \ kpc$ at $z=0.25$ for secondary BHs. This increase in separation between galaxy centre and BH for both main and secondary BH is partially due to the increased size of galaxies at low redshift but not exclusively. Accounting for the increase in average galaxy size, here measured by the galaxies' effective radius $r_{\rm eff}$ with decreasing redshift, the median distance still increases from $1.0 r_{\rm eff}$ at $z=3$ to $1.3 r_{\rm eff}$ at $z=0.25$ for main BHs. We therefore conclude that main BHs struggle to remain attached to the centre of their host galaxy for long periods of time, and both main and secondary BH become less central over time.

\begin{figure}
    \centering
    \includegraphics[width=\columnwidth]{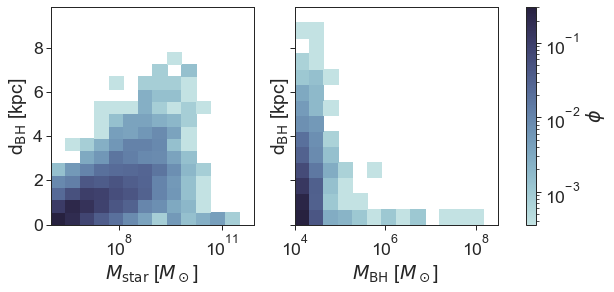}
    \caption{Log-normalised distribution $\phi$ of the distance between main BHs and the centre of their host galaxies, $d_{\rm BH}$ as a function of the host galaxy mass $M_{\rm star}$ (left) and BH mass $M_{\rm BH}$ (right) for a stacked sample of all BHs at $z=3,2,1$ and $0.25$.}
    \label{fig:Mstar_BHpos}
\end{figure}

As can be seen in Fig. \ref{fig:Mstar_BHpos} it is predominantly low-mass BHs in low-mass galaxies which struggle to remain attached to the centre. While BHs close to the centre of their host galaxy exist at all BH and galaxy masses, it is only above a threshold BH mass ($M_{\rm BH} \gtrsim  10^{5} \rm \ M_\odot$) and a threshold host galaxy mass ($M_{\rm star} \gtrsim 10^{10} \rm \ M_\odot$) that main BH can be reliably found in the centre of the galaxy. We will explore the dynamics of both main and secondary BHs in \nh~galaxies across the full range of galaxy masses in detail a companion paper \citep{Beckmann2023}.

From previous work, BHs in galaxies that are not located in the centre of galaxies is not unexpected.  \citet{Bellovary2021} found that BHs in dwarf galaxies are frequently displaced from the centre of their host galaxy following galaxy mergers, and \citet{Sharma2022} report that a significant fraction of their BHs in dwarf galaxies is off-centre. Observationally, there is also  evidence for the fact that BHs in dwarf galaxies wander. \citet{Reines2019} reported that the majority of their radio-selected AGN in dwarf galaxies are off-centre with respect to the host galaxy. Part of this is due to the fact that dwarf galaxies frequently show disturbed morphologies, so there is no clear `centre' for BHs to be located in. However, the displacement of BHs in dwarf galaxies goes beyond this effect, as the fraction of disturbed morphologies in dwarf galaxies falls with decreasing redshift (in \nh, from 20 percent at $z=3$ to 5 percent at $z=1$, see \citealp{Martin2020}), while the mean distance of main and secondary BHs to the centre of the galaxy increases with decreasing redshift.  

By $z=0.25$, only 27 percent of main BHs (and 30 percent of all BHs) remain within 1 kpc of the centre, which is lower than in previous studies of BHs in dwarf galaxies such as \citet{Sharma2022}. This is through a mix of BHs being displaced by galaxy mergers \citep{Bellovary2021} but most likely also a consequence of \nh~lower BH seed mass, as dynamical timescales for BHs to settle back to galactic centres is directly proportional to the mass of the BH  \citep{Pfister2019}, and lower-mass BHs struggle more than higher mass BHs to sink back to the centre \citep{Ma2021,Bellovary2021} (see also Fig. \ref{fig:Mstar_BHpos}). In \nh, BH might additionally be prone to wandering as their seed mass of $M_{\rm BH,0} = 10^4 \rm \ M_\odot$ is close to the stellar mass resolution of $1.3 \times 10^4 \rm \ M_\odot$ of \nh, which induces stochastic effects in their orbits. Additionally, we do not analytically model the unresolved dynamical friction from star and DM particles, which has been shown to play a strong role in allowing BHs to sink to the centre of galaxies \citep{Pfister2019,Chen2021}, and the dynamical friction from gas becomes less efficient at increased resolution due to instabilities in the wake \citep{Beckmann2018} and the turbulent nature of the gas\citep{Lescaudron2022}.

This points to a big open problem in the field: if low-mass BHs struggle to remain attached to the centres of galaxies, and not being attached to the centres of galaxies means BHs struggle to grow, how can we explain the observed, active IMBH in dwarf galaxies? Either there is something that is fundamentally missing in our understanding (or numerical modelling) of BH dynamics, or all seed BHs must be sufficiently massive to avoid such dynamical difficulties. 

\section{Conclusions}
\label{sec:conclusions}

In this paper we studied the evolution of BHs in dwarf galaxies in the \nh~simulation. We found that
\begin{enumerate}
    \item BHs do not start growing efficiently until their host galaxy leaves the dwarf galaxy regime (i.e. when $M_{\rm star} > 3 \times 10^{9} \rm \ M_\odot$. As a result, most BHs in dwarf galaxies remain near their seed mass for long periods of time (here $M_{\rm BH, 0} = 10^4 \rm \ M_\odot$. There is a weak trend for BHs in more massive dwarf galaxies to grow more through accretion than in low-mass dwarfs.
    \item Occupation fractions of BHs in dwarf galaxies remain high and show little evolution with redshift. The fraction of galaxies hosting multiple BHs increases strongly with host galaxy stellar mass and reaches unity as galaxies become massive.
    \item BHs grow much more actively at high redshift ($z\geq2$), where there are a significant number of objects that have time-averaged high Eddingtion ratios. At low redshift ($\leq 1$, average Eddington ratios fall very low but brief bursts of higher activity remain common. These bursts are too infrequent to contribute significant mass growth to the BH, but do make BHs intermittently observable. 
    \item When looking at the X-ray luminosity of BHs and their host galaxies, most BHs are insufficiently luminous to outshine their host even at high redshift. This means that even at high redshift ($z=3$), the fraction of dwarf galaxies that host an AGN that can be detected above an optimistic luminosity threshold of $10^{39} \rm \ erg s^{-1}$ is only $\sim 17 $ percent. 
    \item At lower redshift, the fraction of AGN for a given X-ray luminosity cut decreases, with no observable AGN remaining at $z\sim0.25$. 
    \item Due to their low seed mass, BHs in \nh~struggle to remain attached to the centres of their host galaxies, with the average distance between BHs and galaxy centre increasing from 0.68 kpc at $z=3$ to 2.96 kpc at $z=0.25$. BHs in massive galaxies sink efficiently to the centre.
\end{enumerate}

Overall, the evolution of the BH population in \nh~shows that the lower seed mass exacerbates many of the processes that limit BH growth. Previous simulation work with higher seed mass should consequently be seen as an upper limit to how much BHs in dwarf galaxies can grow given our current model of BH dynamics, BH accretion and SN feedback. As such, dwarf galaxies remain a promising laboratory to constrain stellar feedback and BH physics.

\section*{Acknowledgements}
This work was granted access to the HPC resources of CINES under the allocations c2016047637, A0020407637 and A0070402192 by Genci, KSC-2017-G2-0003 by KISTI, and as a “Grand Challenge” project granted by GENCI on the AMD Rome extension of the Joliot Curie supercomputer at TGCC. This research is part of the Spin(e) ANR-13-BS05-0005 (http://cosmicorigin.org), Segal ANR-19-CE31-0017 (http://secular-evolution.org) and Horizon- UK projects. This work has made use of the Infinity cluster on which the simulation was post-processed, hosted by the Institut d’Astrophysique de Paris. RSB would like to thank Newnham College, Cambridge, for financial support. TK was supported by the National Research Foundation of Korea (NRF) grant funded by the Korea government (No. 2020R1C1C1007079 and No. 2022R1A6A1A03053472). We warmly thank S. Rouberol for running it smoothly. The large data transfer was supported by KREONET which is managed and operated by KISTI. RSB gratefully acknowledges funding from Newnham College, Cambridge. SK acknowledges support from the STFC [ST/S00615X/1] and a Senior Research Fellowship from Worcester College, Oxford.

%%%%%%%%%%%%%%%%%%%%%%%%%%%%%%%%%%%%%%%%%%%%%%%%%%

%%%%%%%%%%%%%%%%%%%% REFERENCES %%%%%%%%%%%%%%%%%%

% The best way to enter references is to use BibTeX:

\bibliographystyle{mnras}
\bibliography{author.bib} % if your bibtex file is called example.bib

% Alternatively you could enter them by hand, like this:
% This method is tedious and prone to error if you have lots of references

%%%%%%%%%%%%%%%%%%%%%%%%%%%%%%%%%%%%%%%%%%%%%%%%%%

%%%%%%%%%%%%%%%%% APPENDICES %%%%%%%%%%%%%%%%%%%%%

%\appendix

%%%%%%%%%%%%%%%%%%%%%%%%%%%%%%%%%%%%%%%%%%%%%%%%%%

% Don't change these lines
\bsp	% typesetting comment
\label{lastpage}
\end{document}